\documentclass{aastex}
\usepackage{ccaption}
\received{3 Sep 2002}
\accepted{15 Oct 2002}

\slugcomment{To appear in the February 10, 2003 issue of Astrophysical Journal}

\begin{document}
\title{BATSE SOFT GAMMA-RAY OBSERVATIONS OF GROJ0422+32}
\author{J. C. Ling }
\affil{Jet Propulsion Laboratory 169-327, California Institute of Technology}
\affil{4800 Oak Grove Drive, Pasadena, CA 91109}
\email{james.c.ling@jpl.nasa.gov}
\and
\author{Wm. A. Wheaton }
\affil{Infrared Processing and Analysis Center, California Institute 
of Technology}
\affil{100-22, Pasadena, CA 91125}
\email{waw@ipac.caltech.edu }

\begin{abstract}

We report results of a comprehensive study of the soft gamma-ray (30 
keV to 1.7 MeV) emission of GROJ0422+32 during its first known 
outburst in 1992. These results were derived from the BATSE 
earth-occultation database with the JPL data analysis package, EBOP 
(Enhanced BATSE Occultation Package).  Results presented here focus 
primarily on the long-term temporal and spectral variability of the 
source emission associated with the outburst, which complement those 
reported earlier by BATSE (Harmon et al. 1993, Ling et al. 2000), 
OSSE (Grove et al. 1998), COMPTEL (Van Dijk et al. 1995) and SIGMA 
(Sunyaev et al. 1993; Roques et al. 1994).

The light curves with 1-day resolution in six broad energy-bands
(e.g. 35-100 keV, 100-200 keV, 200-300 keV, 300-400 keV, 400-700 keV 
and 700-1000 keV) show the high-energy flux ($>$200 keV)
led the low-energy flux ($<$200 keV) by $\sim5$ days in reaching the 
primary peak, but lagged the latter by $\sim7$ days in starting the 
declining phase.
We confirm the  "secondary maximum" of the low-energy ($<$200 keV) 
flux at $\sim$TJD 8970-8981, $\sim120$ days after the first maximum, 
reported earlier by the BATSE team (Harmon et al. 1993).
Our data show that the "secondary maximum" was also prominent in the 
200-300 keV band, but became less pronounced at higher energies.

During this 200-day period, the spectrum evolved from a power-law 
with photon index of 1.75 on TJD 8839, to a shape that can be 
described by a Comptonized model or an exponential power law below 
300 keV, with a variable power-law tail above 300 keV. The spectrum 
remained roughly in this two-component shape until $\sim9$ November 
(TJD 8935) when the 35-429 keV luminosity dropped to below $\sim20$\% of 
its peak value observed on TJD-8848. It then returned to the initial 
power-law shape with an index of $\sim2$ and stayed in this shape until 
the end of the period.  The correlation of the two spectral shapes 
(e.g.Compton/power law tail vs.power law) with the high and low 
luminosities of the soft gamma-ray emission is strongly reminiscent 
of that seen in Cygnus X-1, suggesting that similar processes are at 
work in both systems. We observed also four separate episodes of 
high-energy (400-1000 keV) emission during the first 84 days of the event.
We interpret these results in terms of the 
Advection Dominated Accretion Flow (ADAF) model with possibly a 
"jet-like" region that persistently produced the non-thermal 
power-law gamma rays observed throughout the event.
\end{abstract}


\section{INTRODUCTION}

The $\gamma$-ray source GROJ0422+32 was first discovered by The Burst 
and Transient Source Experiments (BATSE; Fishman et al. 1989) onboard 
the Compton Observatory on August 5th 1992 (Paciesas et al. 1992). 
Early BATSE results (Harmon et al 1992, 1994) reported by the MSFC PI 
team showed that the source underwent a major outburst when the hard 
x-ray (40-230 keV) flux reached a level of $\sim3$ Crab in about three 
days. It remained at this level for approximately three days and then 
decreased exponentially with a time constant of 43.6 days (Vikhlinin 
et al. 1995). A secondary maximum was then observed approximately 120 
days after the first maximum, in December 1992 (Harmon et al. 1994), 
again followed by an exponential decay with time scale similar to 
that of the first maximum.  The entire outburst lasted for about 200 
days before returning to the pre-burst quiescent level.

Observational data in UV/Optical/IR/Radio (Castro-Tirado et al. 
1992, 1993; Wagner et al. 1992; van Paradijs et al. 1994; Shrader et 
al. 1992a, b, 1994; Chevalier and Ilovaisky 1995, 1996; Bonnet-bidaud 
and Mouchet 1995;  Callanan et al. 1996; Casares et al. 1995; Garcia 
et al. 1996) and in X-ray and $\gamma$-ray (Sunyaev et al. 1993; 
Pietsch et al. 1993; Harmon et al. 1994; Roques et al. 1994; 
Vikhlinin et al. 1995; and van Dijk et al. 1995; Grove et al. 1998; 
Ling et al. 2000; Iyudin \& Haberl 2001) of this event and subsequent 
outbursts  have also been reported in the literature. An optical 
counterpart of GROJ0422+32 was observed by Castro-Tirado et al (1992, 
1993) and Wagner et al. (1992) showing a visual magnitude of V = 13 
at the peak of the event. The optical lightcurve then declined 
exponentially with a time scale of 170 days (Shrader et al. 1994) and 
reached a quiescence level of V = 22.35 $\sim800$ days after the 
discovery of the x-ray source (Garcia et al. 1996).  The 9 magnitude 
change in V was the largest seen in any Soft X-ray Transients (SXT) 
to date (van Paradijs \& McClintock 1995).

There was also evidence showing the source is a binary system. 
Filippenko, Matheson, \& Ho (1995) observed a 5.08 $\pm$ 0.01 hour 
orbital period, and estimated a mass function of $f(M) =1.21 \pm 0.06 
M_\odot$ (where $M_\odot$ is the solar mass). The mass function is 
consistent with $f(M) = 0.40 \pm 1.40 M_\odot$ reported by Orosz \& 
Bailyn (1995), and $0.85 \pm 0.30 M_\odot$  reported by Casares et al. 
(1995). Using the orbital inclination of $i \leq 45^\circ$ determined by 
Callanan et al. (1996), the mass of the compact object is estimated 
to be $\geq 3.4 M_\odot$.  However, based on the orbital inclination of 
$10^\circ  - 31^\circ$  estimated independently from infrared and optical 
photometry, a lower limit of $\geq 9 M_\odot$ for the compact object is 
implied, which strongly suggested the existence of a black hole (BH) 
in the system. The distance of the source was estimated by Shrader et 
al (1994) to be $2.4 \pm 0.4$ kpc.

The BH nature of GROJ0422+32 was also supported by X and $\gamma$-ray 
observations. First, the contemporaneous x-ray and soft $\gamma$-ray 
spectra measured by TTM and HXTE instruments (Sunyaev et al. 1993; 
Maisack et al. 1994) in the 2 to 300 keV range, and SIGMA (Roques et 
al. 1994) in the 35-600 keV range, onboard the Mir-Kvant spacecraft, 
and by BATSE (Harmon et al. 1994; Ling et al. 2000), OSSE (Grove et 
al. 1998) and COMPTEL (van Dijk et al. 1995) on the Compton 
Observatory in the 50 keV - 2 MeV range can generally characterized 
by either an exponentially truncated power-law  with a photon 
power-law index of $1.49\pm0.01$, break energy  $E_b$ of $60\pm3$ keV, and 
e-folding energy $E_f$ of $132\pm2$ keV, or in terms of a Comptonized 
disk model (Sunyaev \& Titarchuk 1980) with a temperature of $kT\sim58$
keV and an optical depth of $\sim2$. It is interesting to note that the 
SIGMA spectrum (Roques et al. 1994) shows excess flux from 400 to 600 
keV above the best-fit Comptonized model, and that high-energy flux 
was also observed from 1-2 MeV by COMPTEL (van Dijk et al. 1995). The 
composite spectral shape, Comptonized below and power-law above 400 
keV respectively (see Ling et al. 2000; Figure 3 panels 36-39) 
strongly resembles the standard-state ($\gamma_2$, or x-ray 
low/hard-state) spectrum of Cygnus X-1 (Ling et al.1987, 1997; McConnell et 
al. 2001), the best-known black-hole candidate in our galaxy.

Second, the timing analysis of the hard x-ray data of BATSE (20 - 300 
keV; Kouveliotou et al. 1992, 1993; van der Hooft et al., 1999), and 
SIGMA (45 - 150 keV; Vikhlinin et al. 1995) showed evidence for low 
frequency quasi-periodic oscillation (QPOs) centered at 0.03 and 0.2 
Hz for the former and 0.3 Hz for the latter. These results are 
consistent with the frequencies observed by OSSE (Grove et al. 1994). 
Because of similar low frequency QPO's observed in other black holes 
(BH; Sunyaev et al. 1991; Van der Klis 1994; Kouveliotou 1994), it is 
suggested (Roques et al.1994) that the GROJ0422+32 could be a BH also.

While previous soft $\gamma$-ray results (30 keV to 2 MeV; Harmon et 
al. 1992, 1993, 1994; Sunyaev et al. 1993; Maisack et al. 1994; 
Roques et al. 1994; van Dijk et al. 1995; Grove et al. 1998; Ling et 
al. 2000) have advanced our understanding of the system, information 
concerning the long-term behavior of the source during the outburst, 
however, is far from complete.  The earlier BATSE published 
lightcurve (Paciesas et al. 1992; Harmon et al. 1992, 1993) primarily 
focused on the energy region below 200 keV. The spectra measured by 
COMPTEL, OSSE and SIGMA focused on only isolated periods of the 
200-day event due in part to the relatively spotty coverage of 
pointed observations made by most of these experiments. Questions 
that need to be addressed include (1) how does the source spectrum 
evolve over the course of the event, and (2) what is the long-term 
behavior of the high-energy flux above 200 keV compared to that below 
200 keV (Harmon et al. 1994; Grove et al. 1998)? Answers to these 
questions will shed further light on the mechanism driving the 
outburst, and ultimately the physical makeup of the system itself.

This paper addresses these important questions. Our results were 
obtained using BATSE Earth Occultation data provided by the BATSE PI 
team at MSFC (Fishman et al. 1989), processed and analyzed using the 
JPL Enhanced BATSE Occultation Package (EBOP; Ling et al. 1996, 
2000). A brief description of the EBOP database and technique are 
given in Section 2. Results produced by EBOP are presented in Section 
3.

\section{EBOP Database and Technique}

The BATSE Earth Occultation data were obtained by its eight Large 
Area NaI scintillator Detectors (LADs), each 50.8 cm diameter and 
1.25 cm thick ($\sim2025$ cm$^2$ geometrical area) operating  in the 
energy range 0.02 - 1.8 MeV (Fishman et al. 1989). These eight LADs 
were placed at the eight corners of the spacecraft and provided 
nearly isotropic response to $\sim2/3$ of the sky that not occulted by 
the Earth. They were therefore ideally suited for unprecedented 
sensitive long-term continuous monitoring of gamma-ray sources, using 
the Earth occultation technique (Harmon et al. 1992, 2002; Ling et 
al. 2000). The BATSE Earth Occultation data provided by the MSFC PI 
team consisted of continuous (CONT) data in 16 energy-channels with 
2.048-s resolution for each LAD. These 2.048-s resolution data were 
summed to 16.384-s resolution bins at JPL, and used as input to the 
EBOP analysis system (Ling et al. 1996, 2000). Due to uncertainties 
of the energy edges of the first and the last energy channels, we 
analyzed only the central 14 channels here. The 5274 16-s data bins 
in each day typically yielded $\sim4300$ valid bins after losses and 
gaps. The heart of the EBOP system fitted these $\sim4300$ measured 
count-rates (per day, per energy channel, and per LAD) to a linear 
model in $\sim$45-75 unknown terms that typically included 
contributions from 10-40 point sources (of a total of 64 sources in 
the EBOP input catalog) and $\sim35$ physical background terms. Since 
the source flux in each of the 14 energy bins, and for each LAD, was 
independently determined daily over a very distinct and highly 
variable background for the different LADs, a strength of the BATSE 
experiment and the EBOP 
system is the consistency check they provide on the source fluxes 
measured by the different LADs daily and over the 10-14 day viewing 
periods as shown in Section 3 (see also Wallyn et al. 2001).

\section{Results}
\subsection{Flux Histories}

Figure 1 shows the flux histories, with 1-day resolution, in the 
35-100 keV, 100-200 keV, 200-300 keV, 300-400 keV, 400-700 keV, and 
700-1000 keV energy bands, respectively, covering the period from 27 
June 1992 (TJD 8800) to 23 April 1993 (TJD 9100). Highlights of these 
results include:

a. The 35-200 keV flux rose sharply after the onset of the 
outburst on 5 August 1992 (TJD 8839, labeled "1"), and reached the 
first of two maxima during the peak of the outburst on 14 August (TJD 
8848, vertical dashed line "b"). It then decreased slightly ($\sim5$\% 
for 35-100 keV, and $\sim6$\% for 100-200 keV) over the next few days 
before rising again to a second maximum on 21 August (TJD 8855, 
vertical dashed line "c"). The declining phase of the event took 
$\sim180$ days. The combined 35-200 keV flux reached half the peak 
level on $\sim$TJD 8877, and a tenth of the peak on $\sim$TJD 8935, 
approximately 80 days later. It continued to decrease to a level 
about 4\% of the peak value on TJD 8960 before slowly rising again to 
the so-called "secondary maximum" on TJD 8972 ($\sim120$ days after the 
primary maximum on $\sim$TJD 8852) at a level about 15\% of the peak 
value. The 35-200 keV flux stayed within 20\% of the "secondary 
maximum" for $\sim20$ days before declining, and finally reached the 
"pre-outburst" quiescent level on TJD 9040.

b. We observed also for the first time energy-dependent 
flux variability above 200 keV (see Figure 1 panels 3-6). These data 
show temporal features significantly different than those observed 
below 200 keV. First, the high-energy flux rose more promptly after 
the onset and reached the 1st of four local maxima on TJD 8843 
(vertical dashed line "a"), 5 days before the 1st maximum of the 
35-200 keV flux (vertical dashed line "b"). It then declined in the 
next nine days before rising to a second maximum on 28 August (TJD 
8862; vertical dashed line in panel "d"), 7 days after the second 
maximum (vertical dashed line in panel "c") and during the declining 
phase of the hard x-ray (35-200 keV) flux. The flux ratio of the 2nd 
maximum to 1st maximum increased with energies, ranging from $\sim1$ 
for 200-300 keV, 1.3 for 300-400 keV, $\sim2$ for 400-700 keV to 
$\sim3.3$ for 700-1000 keV. The local minimum around TJD 8850 became 
more pronounced with increasing energy. The flux ratio of this 
minimum to the 2nd maximum decreased from $\sim0.9$ for 200-300 keV, 
$\sim0.4$ for 300-400 keV, $\sim0.2$ for 400-700 keV to $\sim0$ for 
700-1000 keV. Variable high-energy fluxes were also consistently 
observed in the three high-energy channels (300-400 keV, 400-700 keV, 
and 700-1000 keV) in at least two other periods centered at $\sim$TJD 
8889 (3rd maximum, vertical dashed line "e") and TJD 8919 (4th 
maximum, vertical dashed line "f"), interspersed with periods of low 
gamma-ray flux at $\sim$\-TJD 8875 and TJD 8900, respectively. Spectra 
measured during these highly variable gamma-ray flux periods are 
presented in detail in the next section. The broad "secondary 
maximum" observed in 35-200 keV was also prominently observed in 
200-300 keV at $\sim$TJD 8970-8981, $\sim120$ days after the primary 
peak, but became less pronounced at higher energies.  There is also a 
dip in intensity at around TJDs 8973-8977 in the midst of the 
"secondary maximum", similar to that seen in the primary peak. Such a 
dip is consistently seen in all six energy-bands. Its significance is 
estimated to be $\sim5\sigma$ at 200-300 keV. Iyudin \& Haberl (2001) 
showed that the gamma-ray emission, based on 25 observations by 
COMPTEL obtained between August 1992 and August 1997 ($\sim1800$ days) 
and three Viewing Periods before the 1992 outburst, is primarily 
confined between 1.5 and 2 MeV, and was more prominent during phases 
from 0.0 to 0.5 of the 120d period, where zero phase corresponds to 
TJD 8840.5. Our BATSE results show that the high-energy fluxes were 
confined to phases from 0.0 to 0.7 of the same periodicity.  The 
averaged flux of the 700-1000 keV emission integrated over TJD 
8841-8923 (phase $\sim$0-0.7) is  $(1.6 \pm 0.3) \cdot 10^{-6}$
photons cm$^{-2}$-s$^{-1}$-keV$^{-1}$ compared to $(-4.9 \pm 5.1) \cdot 10^{-7}$ 
photons cm$^{-2}$-s$^{-1}$-keV$^{-1}$ integrated over TJD 8925-8960 ($\sim$phase 
0.7-1.0). Similarly for the 400-700 keV emission, the averaged fluxes 
for the same two periods are $(10.0 \pm 0.6) \cdot 10^{-6} $
photons-cm$^{-2}$-s$^{-1}$-keV$^{-1}$ and $(1.3 \pm 0.8) \cdot 10^{-6} $
photons-cm$^{-2}$-s$^{-1}$-keV$^{-1}$, respectively.

\subsection{Spectra}

The complex energy-dependent flux histories shown in Figure 1 imply 
complex spectral changes over the course of the 200-day event. We 
selected a sample of thirty-six single-day spectra spanning the 
period from TJD 8839 to TJD 9033 to show such changes. Pertinent 
information related to each of these spectra is given in both Figure 
2 and Table 1. For each day, the source was observed by two to four 
of the eight BATSE LADs with good sensitivity. These "source-viewing" 
LADs, which are identified in Table 1 and Figure 2, were selected 
using the criterion discussed by Ling et al. (1996, 2000). Each LAD 
spectrum has 14 energy bins. The solid line is the best-fit model 
(either Compton model or power-law) to the $n$ data points, where $n$ = 
number of LADs x 14 energy channels, using the standard analysis 
fitting program XSPEC (Arnaud 1996).  The best-fit model and 
parameters, as well as reduced $\chi^2$ ($\chi^2$/$\nu$, where $\nu$
is the number of degrees of freedom) of the fit are 
also displayed in each panel and 
listed in Table 1. In twelve of the thirty-six panels (e.g. panels in 
the middle column in each of the four pages) in Figure 2, spectra 
measured simultaneously by all "source-viewing" LADs on that day are 
shown and compared. Since each LAD-spectrum was independently 
determined over a complex background that is totally different from 
that of the other seven LADs, these simultaneously measured spectra 
allow one to assess the quality and consistency of the results. 
Consistency of fluxes among all "source-viewing" LADs is also 
reflected by the goodness of the model fits to the data (Ling et al. 
1996; Ling et al. 2000). Large reduced $\chi^2$ could be caused by either 
inconsistency of LAD fluxes or inadequacy of the spectral model. 
Twenty-nine of the 36 spectra shown in Figure 1 have acceptable fits 
with either a power law or a Compton model. The relatively large 
reduced $\chi^{2}$ shown in panels 13-21 ($\sim$TJD 8855-8883, taken 
around 
the 2nd maximum of the high-energy fluxes) are due primarily to a 
spectral tail above 300 keV.  In general the consistency among LADs 
in the 36 spectra in Figure 2 gives us good confidence in the 
analysis technique and results. Having developed data analysis 
systems for four separate missions (e.g. OSO-7, HEAO-1, HEAO-3, and 
CGRO) in the past three decades, and shared the frustration of many 
others of the disparate results produced by different experiments, 
some of which could be caused by systematic effects associated with 
the inadequacy of the data analysis techniques, we believe it is 
useful to show internal consistency achieved among the relevant 
detectors of a single instrument, such as BATSE.  This was an 
important goal for EBOP. We hope the results presented here will 
encourage other investigators to show internal consistency, as a 
necessary quality control requirement in future gamma-ray data analysis

Key spectral results shown in Figures 2 \& 3 and Table 1 are 
summarized as follows:

\subsubsection{High-State Spectrum and Variability}

a.	During the rising phase of the event, the single-day spectrum 
shown in Figure 2a changed from a power-law with photon index of 1.75 
on TJD 8839 (Figure 2a panel 1) to two days later, a shape better 
characterized by a single temperature analytic Compton model (Sunyaev 
\& Titarchuk 1980), with kT = 59$\pm5$ keV and $\tau$ = 2.27$\pm0.19$,
on TJD 8841 (Figure 2a-panel 2). The reduced $\chi^2$ for a 
power-law fit to the TJD 8841 spectrum is 2.9 compared to 0.66 
for the Compton model (see 
Table 1), clearly indicating the change in spectral shape.  In fact, 
the 4-day averaged spectra measured by the two "source-viewing" LADs 
(1 $\&$ 5) during Viewing-Period 35 (VP-35 TJD 8841-8844; see Figure 3a 
- panel 1) show excesses of the high-energy ($>$300 keV) flux over the 
best-fit Compton model (solid line -kT =51.1 keV, $\tau$ = 2.93). The 
dashed line is the best-fit power law ($\alpha$ = 3.4) to the flux in the 
five high-energy channels (313 - 1700 keV) measured by the two 
"source-viewing" LADs (1 $\&$ 5). Such high-energy flux was only hinted 
in the corresponding single-day spectra (Figure 2a panels 2-5) due to 
statistical limitation of the data. The two-component (Compton + 
power law) spectrum is strongly reminiscent of the $\gamma_2$ spectrum (or 
"low/hard" X-ray state) of Cygnus X-1 (Ling et al. 1997; McConnell et 
al. 2000; Ling 2001), suggesting similar physical processes were at 
work in both systems.

b.	Figure 2-a,b,c panels 6-25 and Figure 3-a,b panels 2-8 show 
the single-day and multiple-day VP spectra, respectively, covering 
the period from TJD 8847 to 8925. This period includes:  (1) the peak 
and decay phases of the event when the 35-429 keV luminosity was 
above 20$\%$ of the peak value (see Table 1), and (2) periods of the 2nd 
(vertical dashed line "e") and 3rd (vertical dashed line "f") 
high-energy  peaks described in Section 3.1. During this 78-day 
period, the spectrum underwent significant changes.  For the 
low-energy Comptonized component below 300 keV, as reflected by the 
best-fit parameters of the single-day spectrum shown in Table 1, the 
electron temperature (kT) varied from 40 to 60 keV and optical depth 
$\tau$ from 2.3 to 3.4. For the high-energy power-law component above 300 
keV, seen more prominently in the VP spectra shown in Figure 3-a,b 
(VPs 36.7, 37, 38, 39, 40, 41 and 42), the best-fit photon index 
varied from $\sim1$ in VP-42, to $\sim3.8$ in VP-39. Details shown in 
Figure 3-a,b also include:

$\bullet$	Panel 2 shows the weighted averaged spectrum measured by four 
BATSE LADs (LADs 0, 2 4 \& 6) during VP-36.5 (TJD 8847-8853), the 1st 
local high-energy flux minimum described in Section 3.1-B (see also 
Figure 1).  The solid line is the best-fit Compton model with kT = 
42.2 keV and $\tau$ = 3.52 to the four "source-viewing" 
LAD-spectra (LADs 0, 2, 4, $\&$ 6).  
The reduced $\chi^2$ for the fit is 1.38 for 53 degrees of 
freedom. Shown also in this panel is a comparison of the BATSE 
spectrum with those measured simultaneously by OSSE (retrieved from 
the OSSE archive in 1995; see also Ling et al. 2000) and COMPTEL (van 
Dijk et al. 1995). Below 429 keV, OSSE fluxes were generally lower 
than those of BATSE by $\sim$5-25$\%$. Above 429 keV, upper limits 
measured by BATSE are consistent with the positive fluxes in the 
518-678 keV bin measured by OSSE, and in the 750-1000 keV and 
1000-2000 keV bins measured by COMPTEL (van Dijk et al 1995). The 
high-energy fluxes measured by COMPTEL and OSSE in this period were 
lower than those measured by BATSE in VP-35 (e.g.by approximately a 
factor of two at 1 MeV. They provide, however, an important 
confirmation of the high-energy component of the two-component 
(Compton and power law) spectrum observed by BATSE in VP-35, and 
further suggest that such component is variable.

$\bullet$	The high-energy component $>$300 keV was also clearly visible 
in the four VP spectra that follow (Figure 3-a,b-panels 3-6: VP-37, 
38, 39 and 40) covering the period from TJD 8855 to TJD 8902. The 
power-law indices for these four spectra vary from 2.7 to 3.8. The 
two larger indices of 3.44 and 3.89 for VP-37 and VP-39, 
respectively, correspond to the 1st (between "a" and "d") and 2nd 
(between "d" and "e") minimun of the high-energy flux shown in Figure 1 
panels 5 \& 6. None of these VP spectra can be adequately fitted with 
a single-component Compton model (solid line). The reduced $\chi^2$ for 
fitting these spectra varies from $\sim2.5$ to $\sim8.4$. The reasons for 
the poor fit are: (1) the presence of the high-energy flux above 300 
keV in the spectrum, and (2) short-term (single-day) spectral changes 
shown in Figure 2, due possibly to intrinsic changes in the system, 
preclude any possibility for a simple model (e.g. Compton model) to 
adequately fit the long-term VP spectrum.  For three of the four VP 
spectra (VP-37, 38 and 40) in which high-energy fluxes are 
prominently visible, data measured by all "source-viewing" LADs were 
display for demonstrating their consistency. The dashed line in each 
panel represents the best-fit power law to the fluxes measured by all 
"source-viewing" LADs in the five high-energy channels (313 - 1700 
keV). Differences in spectral indices among the VP spectra reflect 
variability in the high-energy fluxes observed during this period. A 
relatively hard high-energy component was seen in the VP-38 and VP-40 
spectra with indices of 2.65 and 2.87, respectively. They correspond 
to the 2nd and 3rd sub-maxima shown in panels 5 $\&$ 6 in Figure 1. 
Included in Figure 3-a,b panels 3 $\&$ 5 are also simultaneous OSSE 
VP-37 and VP-39 spectra, respectively, extracted from the OSSE 
unpublished archive in 1995 (Ling et al. 2000). Again, the OSSE 
fluxes are systematically lower than those of BATSE below 500 keV by 
few to 26$\%$ in VP-39, and 25-47$\%$ in VP-37. More significantly, the 
weighted averaged VP-37 fluxes in energy bins of 766-1104 keV and 
1104-1700 keV of (3.35 $\pm$ 1.82) $\cdot$ 10$^{-6}$ and 
(1.52 $\pm$ 0.81) $\cdot$ 10$^{-6}$
photons/cm$^{2}$-s-keV measured by BATSE are higher than the 
OSSE 3$\sigma$ upper limits of $\sim9 \cdot 10^{-7}$ and $8 \cdot 10^{-7}$
in energy bins 887 - 1160 keV and 
1160-1517 keV, respectively. The reason for the discrepancy is not 
known at this time. Since individual OSSE VP spectrum has not been 
published in open literature, such discrepancyare therefore not 
formally confirmed. Finally, no high-energy flux was observed 
inVP-41, which corresponds to the 3rd minimum (between "e" and "f") 
shown in Figure 1 panel 5 $\&$ 6.

$\bullet$	The hardest high-energy component was seen in the VP-42 
spectrum with a power-law index of $\sim1$. Positive fluxes were 
measured in each of the five high-energy channels  (313-429 keV, 
429-595 keV, 595-766 keV, 766-1104 keV and 1104-1700 keV) at 13.8, 
7.9, 6.2, 5.7, and 6.7$\sigma$ significance, respectively. Furthermore, they 
were consistently seen by the two "source-viewing" LADs  (5 $\&$ 7). The 
difference of the fluxes in each energy band measured by the two LADs 
was estimated to be 1.1, 0.8, 2.0, 0.4, and 0.2$\sigma$ significance.
Such hard MeV component reminds us of the MeV "bump" seen 
in the $\gamma_{1}$ 
spectrum of Cygnus X-1 (Ling et al. 1987).

c.	Figure 4 shows a comparison of spectra measured 
simultaneously by BATSE and SIGMA (Roques et al 1994) during the 
period from TJD 8850 to TJD 8891. Above 300 keV, BATSE results were 
consistent with those of SIGMA. Below 300 keV, SIGMA results were 
generally lower by $\sim20\%$ at 50 keV, $\sim33\%$ at 100 keV 
and $\sim43\%$ 
at 200 keV. Since the source fluxes were highly variable during this 
40-day period, part of the discrepancy below 300 keV may be due to 
differences in weighting the short-term daily fluxes in deriving the 
averaged spectrum.

\subsubsection{Low-State Spectrum}

When the 35-500 keV luminosity dropped below to 20$\%$ of the peak value 
observed on TJD 8848 (see Table 1) after TJD 8925, the source 
spectrum returned to a power law with indices of $\sim1.8$-2 (see 
Figure 2 panels 28-36), similar to that first seen at the start of 
the event on TJD 8839. Figure 5 shows a comparison of a 30-day 
averaged low-state spectrum observed on TJD 9010-9040 with the 
average high-state spectrum covering the six periods shown in Figure 
3 from TJD 8841 to TJD 8865 (VPs 35-38). The two spectra are clearly 
distinct. Above 300 keV, the low-state spectrum with a power index of 
$\sim2$ is harder than that of the high-state spectrum with index of 
$\sim5.2$, and intersects the latter at $\sim600$ keV. Similar spectral 
features and characteristic were also observed in Cygnus X-1 between 
the x-ray low-state and high-state spectra (McConnell et al. 2002).

\section{Discussion}

A comprehensive study of the long-term spectral and temporal 
properties of soft gamma-ray emission of GROJ0422+32 shown in the 
1992 outburst is the primary subject of this report. Highlights of 
our results are as follows:

$\bullet$ The light curves in the six energy bands (Figure 1) show that the 
high-energy ($>$200keV) flux led the low-energy flux ($<$200 keV) by 
$\sim5$ days in reaching the initial peak, but lagged the latter by 
$\sim7$ days before starting to decline.

$\bullet$ We confirm the secondary maximum in the low-energy ($<$200 keV) flux 
at $\sim$TJD 8970-8981, $\sim120$ days after the first maximum, as 
reported earlier by the BATSE team (Harmon et al. 1993). Such a 
secondary maximum was also prominently observed in the 200-300 keV 
band, but became less pronounced at higher energies (Figure 1). We 
also observed a dip in intensity in all six energy-bands at $\sim$TJDs 
8973-8977 in the midst of the secondary maximum.  The dip is similar 
to that seen during the primary peak $\sim120$ days earlier.

$\bullet$ During this 200-day period, the spectrum evolved from the low 
intensity power-law shape with photon index of 1.75 on TJD 8839, to a 
high-intensity shape of two components: a themal-Comptonization shape 
below 300 keV, with a power-law tail above 300 keV with variable 
index from $\sim1$ to $\sim4$ (Figures 2 $\&$ 3).

$\bullet$ The spectrum remained roughly in this two-component shape until 
$\sim9$ November (TJD 8935) when the 35-429 keV luminosity dropped to 
$\sim20\%$ of its peak value observed on TJD-8848. At that time, the 
spectrum returned to the initial power-law shape with an index of 
$\sim2$ and stayed in this shape until the end of the event (Figures 2 
$\&$ 3; Table 1).

$\bullet$ Strong episodes of high-energy (400-1000 keV) emission were 
observed on four separate occasions during the first 84 days of the 
event (Figure 1 panels 5-6). This corresponds to 0-0.70 phase of the 
120-day period, which was first suggested by Iyudin $\&$ Haberl (2001) 
based on several yeas of COMPTEL observations. COMPTEL results showed 
that the 1.5-2 MeV emission was primarily confined to 0-0.5 phase of 
the 120-day period

$\bullet$ The averaged high-intensity spectrum above 300 keV obtained on TJD 
8841-8865 is softer than the average low-intensity power-law spectrum 
(TJD 9010-9040), and intercepts the latter at $\sim600$ keV (Figure 5).

$\bullet$ Several key features displayed by GROJ0422+32 spectra are 
remarkably similar to those seen in Cygnus X-1suggesting that similar 
processes may be at work in both systems. A direct comparison of 
these features is shown in Table 2:

The two-component Compton plus power-law spectrum observed in 
GROJ0422+32, Cygnus X-1 and several other BH binaries has been the 
subject of several theoretical studies in recent years. Earlier works 
in interpreting the Cygnus X-1 spectra below $\sim1$ MeV in terms of a 
two-region core/corona thermal model (Ling et al. 1997; Skibo \& 
Dermer1994) using the Monte Carlo approach, have had some degree of 
success. However, we have now seen persistent power-law emission 
extended to  $\sim1$ MeV in both the high and low-intensity 
$\gamma$-ray state 
spectra of GROJ0422+32 as shown in this and other papers (Grove et 
al. 1998; Van Dijk et al. 1995; Iyudin \& Habert, 2001), and in the
$\gamma_{0}$ 
(high/soft) and $\gamma_{2}$ (low/hard) spectra of Cygnus X-1 (Ling et al. 
1997; McConnell et al. 2000, 2002). These results suggest that the 
high-energy power-law component cannot be adequately interpreted in 
terms of pure thermal processes alone, and that non-thermal processes 
must be also at work in these systems.
Power-law spectral tail may be associated with dynamical 
Comptonization processes in converging flows onto a black hole 
(Laurent \& Titarchuk 1999; Turolla et al. 2002). Turolla et al. 
(2002) showed that a power-law photon index of $<$3 can be produced by 
up-scattering of primary photons off in-falling electrons. However, 
no direct comparison between theoretical predictions and 
observational data can be obtained at this time. Non-thermal 
gamma-ray emission may be also associated with jets, which was 
discussed by Meier (2001) as a natural consequence of accretion flows 
onto rotating black holes. A relativistic jet in Cygnus X-1 has been 
observed in the radio band (Stirling et al. 1996; Fender et al. 2000; 
Fender 2001) when the source was in the $\gamma_{2}$ (low/hard) state.
Radio 
emission was also seen in GROJ0422+32 (Shrader et al. 1994), although 
no jet-like structure was resolved from these observations.
  A more general approach using a hybrid thermal/non-thermal 
comptonization model  (EQPAIR) was proposed by Coppi (1998) (see also 
Gierlinski et al., 1999). In this model, the electron distribution 
consists of a Maxwellian component with a temperature, kT, plus a 
non-thermal power-law component.  The acceleration of non-thermal 
electrons is independently taking place but is coupled to the 
background thermal plasma by Compton scattering and Coulomb collision 
processes. The model basically allows for both thermal and nonthermal 
comptonization of soft photons, as well as pair production, Compton 
reflection, and bremsstrahlung emission.
Although we do not know with certainty at the present the processes 
for producing the non-thermal power-law $\gamma$-ray emission 
(e.g. jets or 
other processes), our data suggest that non-thermal power law 
emission was present throughout the outburst. It was fully visible in 
the 35 keV to 1 MeV energy band when the GROJ0422+32 was in the 
low-intensity state, and only partially visible in the 313 - 1 MeV 
band when it was in the high-intensity state. Furthermore, the 
average spectral index for the former was harder ($\sim2$) than the 
latter ($\sim3$-5).
We suggest a possible scenario for interpreting the observed data 
that includes a separate non-thermal (perhaps a jet-like) source 
region in the ADAF model of Esin et al (1998) along with the source 
geometry envisioned by Poutanen \& Coppi (1998) and others (Coppi 
1998; Fender \& Kuulkers 2001; Zdziarski 2002) (see Figure 6). In this 
scenario, during the high-intensity state (or $\gamma_{2}$ state for Cygnus 
X-1; Figure 6 right panel), the system consists of a hot inner 
corona, a cooler outer thin disk, and a region that produced the 
variable power-law $\gamma$-ray emission. Under this condition, the 
transition radius of the disk is $\sim100$ Schwarzschild radii from the 
black hole. Electrons in the hot corona up-scattered the low-energy 
photons produced both inside the corona as well as from the outer 
disk to form the Comptonized component that dominates the spectrum in 
the 35-300 keV range. They also down-scattered the high energy 
photons ($>$10 MeV) produced in the "jet" region resulted in forming a 
softer power-law component observed in the 300 keV to 1 MeV range 
compared to that observed in the low-intensity spectrum.


When the source was in the low-intensity soft $\gamma$-ray state 
(or $\gamma_{0}$ 
state for Cygnus X-1) due probably to a significantly increase of the 
accretion rate, a large soft flux was produced in the disk that 
effectively quench and cool the inner corona, and moved the 
transition radius inward to a distance very close to the horizon 
(Figure 6 left panel). Under this condition, the Comptonized 
component in the 35-300 keV range diminishes, and the source spectrum 
is dominated by the unperturbed power-law emission produced in the 
"jet"-like non-thermal source region with a characteristic  index of 
$\sim2$.

While the above scenario helps to interpret some aspects of the new 
observational features, there are several issues that are not yet 
resolved: (1) what is the cause for the time-lag effect of $\sim$few 
days seen between the low ($<$200 keV) and high-energy ($>$200 keV) 
photons? Is this an indication of the response time for the 
Comptonization process in the system to a sudden change of the 
accretion rate? (2) The hard high-energy (0.3-1 MeV) spectral 
component observed in VP-42 with an index of $\sim1$ is strongly 
reminiscent of the "MeV bump" seen in the $\gamma_{1}$ 
spectrum of Cygnus X-1 
(Ling et al. 1987). Is this caused by a intrinsic change in the 
non-thermal emission, or is it a signature of the pair plasma (Liang 
and Dermer 1988; Ling $\&$ Wheaton 1989; Poutanen $\&$ Coppi, 1998) 
produced by the heating the corona to very high temperature ($\sim10^{9}$\- 
K). Such heating could be caused by a further reduction of the accretion 
rate that led to the reduction of soft photons for its cooling?  (3) 
The intensity of the radio emission was observed to track the gamma 
rays throughout the outburst for GROJ0422+32 (Shrader et al. 1994), 
but was only seen in the $\gamma_{2}$ state (and not the 
$\gamma_{0}$ state) of Cygnus 
X-1 (Stirling et al. 1996; Fender et al. 2000; Fender 2001). What is 
the reason for this difference?  Is it caused by a difference of the 
"beaming" effect in the two systems?  We hope these results will 
stimulate further theoretical and observational investigations in the 
future of this very unusual and exciting black-hole system
discovered by BATSE.

\acknowledgements
	We wish to thank Gerald Fishman, Alan Harmon, Mark Finger and 
Geoff Pendleton of the BATSE team for their support of the BATSE 
Earth Occultation investigation effort at JPL through-out the years, 
Dave Meier and Paolo Coppi for their extremely useful comments on the 
interpretation of these data, and undergraduate students Robert Kern, 
Zachary Medin, and Juan Estrella for processing the data. The work 
described in this paper was carried out by the Jet Propulsion 
Laboratory, under the contract with the National Aeronautics and 
Space Administration.

\newpage
FIGURE CAPTIONS

\figcaption{Flux histories, with 1-day resolution, in six separate 
energy-bands are shown. The high-energy (see panels 3-6) flux rose 
more sharply and reached the first of the four episodic peaks  (shown 
by four vertical dashed lines "a", "d", "e" and "f") on TJD 8843 
("a") five days before the first low-energy sub-peak (vertical dashed 
line "b")  on TJD 8848. The decrease of the low-energy flux between 
the first and the second sub-peaks on TJD 8855 (vertical dashed line 
"c") is only $\sim4\%$.  However, the decrease of the high-energy fluxes 
(3rd to 6th panels) between its first and second sub-peak on TJD 8862 
(vertical dashed line "d") is significantly more pronounced. The 
high-energy fluxes also lagged the low-energy fluxes in starting the 
decline phase by $\sim7$ days (see vertical dashed lines "c" and "d"). 
A broad "secondary maximum" for the low-energy fluxes $\sim120$ days 
later at $\sim$TJD 8970-8981 which was observed and reported by the 
BATSE team earlier (Harmon et al. 1993) was also prominently observed 
in the 200-300 keV bin, but became less pronounced at higher 
energies. Note that there is also evidence for a "dip" in intensity 
at around TJD 8977 in the midst of the "secondary maximum" which is 
shown in all six energy bands.}

\figcaption{(a) A sample of 9 single-day spectra was selected to show 
the evolutionary changes ofthe source spectrum during the first 
twelve days (TJD 8839-8850)of the 200-day event (see also Figure 1). 
The spectrum evolved from the "low-intensity" power-law shape at the 
onset on TJD 8839 to a "high-intensity" Comptonized shape $<$300 keV 
two days later on TJD 8841. A high-energy power-law component $>$300 
keV was also observed prominently in the VP-35 spectra shown in 
Figure 3a during the first of the four "episodic" 
$\gamma$-ray emission 
periods. However, it was not clearly visible in the single-day 
spectra due to statistical limitation of the data. Note that in three 
panels of the middle column, spectra measured simultaneously by all 
"source viewing" LADs on that day are shown and compared. This is a 
self imposed consistent test to serve as "quality control" to ensure 
credibility of all EBOP results. (b) A second set of 9 single-day 
spectra covering the period TJD 8851-8862 (identified as "10" to 
"18"in Figure 1) During this period, the fluxes, specifically those 
$>$200 keV, gradually rose from a local minimum at $\sim$TJD 8850 to a 
peak at $\sim$TJD 8862. While the spectrum averaged over the entire 
period (VP-37, see Figure 3c) remained in the same two-component 
shape,  namely a Comptonized component $<$300 keV followed by a 
variable power-law $>$300 keV that was visible even in the single-day 
spectra on TJD 8858, 8859 and 8862 (panels 16-18), respectively. (c) 
A third set of 9 single-day spectra covering the period TJD 8863-8945 
(identified as "19" to "27"in Figure 1).  During this period, the 
spectrum remained approximately in the same two-component shape  (see 
also Figure 3-a,b panels 4-8), namely a Comptonized component below 
300 keV and a variable power-law $>$300 keV, until $\sim$TJD 8935 when 
the 35-429 keV luminosity dropped to $\sim20\%$ of the peak value (see 
also Table 1). The spectrum then returned to the initial power law 
shape as that shown on TJD 8839 (Figure 2a panel 1) with index of 
$\sim2$ (see panels 26 \& 27). (d) A fourth set of 9 single-day spectra 
covering the period TJD 8955-9033(identified as "28" to "36"in Figure 
1). The spectrum stayed in the power law shape throughout this period 
until the end of the event.}

\figcaption{(a) Four consecutive VP (Viewing Periods 35, 36.5, 37 and 
38) spectra covering the initial phases of the outburst from TJD 8841 
to TJD 8865. A variable high-energy ($>$300 keV) power-law component 
superposed to the Compton component below 300 keV, is clearly visible 
in panels 1, 3, and 4 that correspond to the two of the high-flux 
periods ("a" and "d") shown in Figure 1. High-energy ($>$400 keV) 
fluxes measured by COMPTEL (van Dijk et al. 1995) and OSSE (from the 
OSSE archive, see Ling et al. 2000) during VP-36.5 are also included 
in panel 2 for comparison. Shown also in panel 3 for comparison are 
unpublished OSSE spectra for VPs 37.  OSSE fluxes below 500 keV are 
generally lower than those of BATSE by $\sim25-47\%$ for VP-37.  (b) A 
second group of consecutive VP (Viewing Periods 39, 40, 41 and 42) 
spectra covering the period from TJD 8867 to TJD 8923. During this 
period, two strong episodes of gamma-ray emission were seen that 
peaked at $\sim$TJD 8889 (see Figure 1 "e")  and TJD 8919 (see Figure 1 
"f"), respectively. A strong variable high-energy ($>$300 keV) spectral 
component superposed to the Compton component below 300 keV, is 
clearly visible in the corresponding VP-40 and VP-42 spectra. OSSE 
unpublished spectrum for VP- 39 (from OSSE archive, see Ling et al. 
2000) respectively is also included in panel 7 for comparison. They 
are generally lower than those of BATSE by few-26$\%$. The strong 
400-1700 keV fluxes that were consistently measured by the two 
"source viewing" LADs (5 \& 7) in VP-42 with a power-law index of 
$\sim1$, is reminiscent of the MeV "spectral bump" seen in the $\gamma_{1}$ 
spectrum of Cygnus X-1 (Ling et al. 1987)}.

\figcaption{A direct comparison of the spectra measured 
simultaneously by BATSE and SIGMA (Roques et al.1994; E. Jourdain 
2002, private communication) during a $\sim40$-day period between TJD 
8850 and 8891. Above 300 keV, BATSE results were consistent with those of SIGMA. Below 300 keV, SIGMA results were generally 
lower by $\sim20$\% at 50 keV, $\sim33$\% at 100 keV and $\sim43$\% at 200 
keV. The discrepancy may be due to differences in weighing the 
short-term variable fluxes in the long-term averages.}

\figcaption{Averaged high-intensity spectrum (TJD 8841-8865) and 
low-intensity spectrum (TJD 9010-9040) have distinct shape. The 
former has two components: a Comptonized component below 300 keV 
followed by a power law (dashed line) above 300 keV, while the latter 
has only one, a power law. The two spectra intersect at $\sim600$ keV. 
These features are similar to those seen in Cygnus X-1 (Ling et al. 
1997; Phlips et al, 1996; McConnell et al. 2001; McConnell et al 
2002) for the $\gamma_{2}$ and $\gamma_{0}$-state spectra which intercepted at $\sim1$ MeV. 
Such strong resemblance of the spectral properties of the two sources 
strongly suggests that similar processes were at work in both 
systems.}

\figcaption{A simple sketch of the system by including a "jet-like" 
region that produced the non-thermal gamma-ray emission in the ADAF 
model of Esin et al (1998) along with the source geometry envisioned 
by Poutanen \& Coppi (1998) and others (Coppi 1998; Fender \& Kuulkers 
2001; Zdziarski 2002). During the "high-intensity" state (right 
panel), the system consists of a hot inner corona, a cooler outer 
thin disk, and a jet-like region that produced the variable power-law 
$\gamma$-ray emission. Under this condition, the transition radius of 
the disk is $\sim100$ Schwarzschild radii from black hole. Electrons in 
hot corona up-scattered the low-energy photons produced both inside 
the corona as well as from the outer disk to form the Comptonized 
component observed in the 30-300 keV range. The same electrons also 
down-scattered the high energy power-law photons produced in the 
"jet" region resulted in forming a softer power-law component 
observed in the 300 keV to 1 MeV range compared to that observed in 
the "low-intensity" spectrum. During the "low-intensity" state due 
probably to a significantly increase of the accretion rate, a large 
amount of soft photons produced in the disk effectively cool and 
quench the corona, and moved the transition radius inward to a 
distance very close to the horizon (Figure 6 left panel). Under this 
condition, the Comptonized component in the 30-200 keV range 
diminishes, and the soft $\gamma$-ray spectrum is therefore dominated by the 
unperturbed non-thermal emission in the "jet-like" region with a 
characteristic power law index of $\sim2$.}

\begin{figure}[t]
\centering
\includegraphics[scale=0.950]{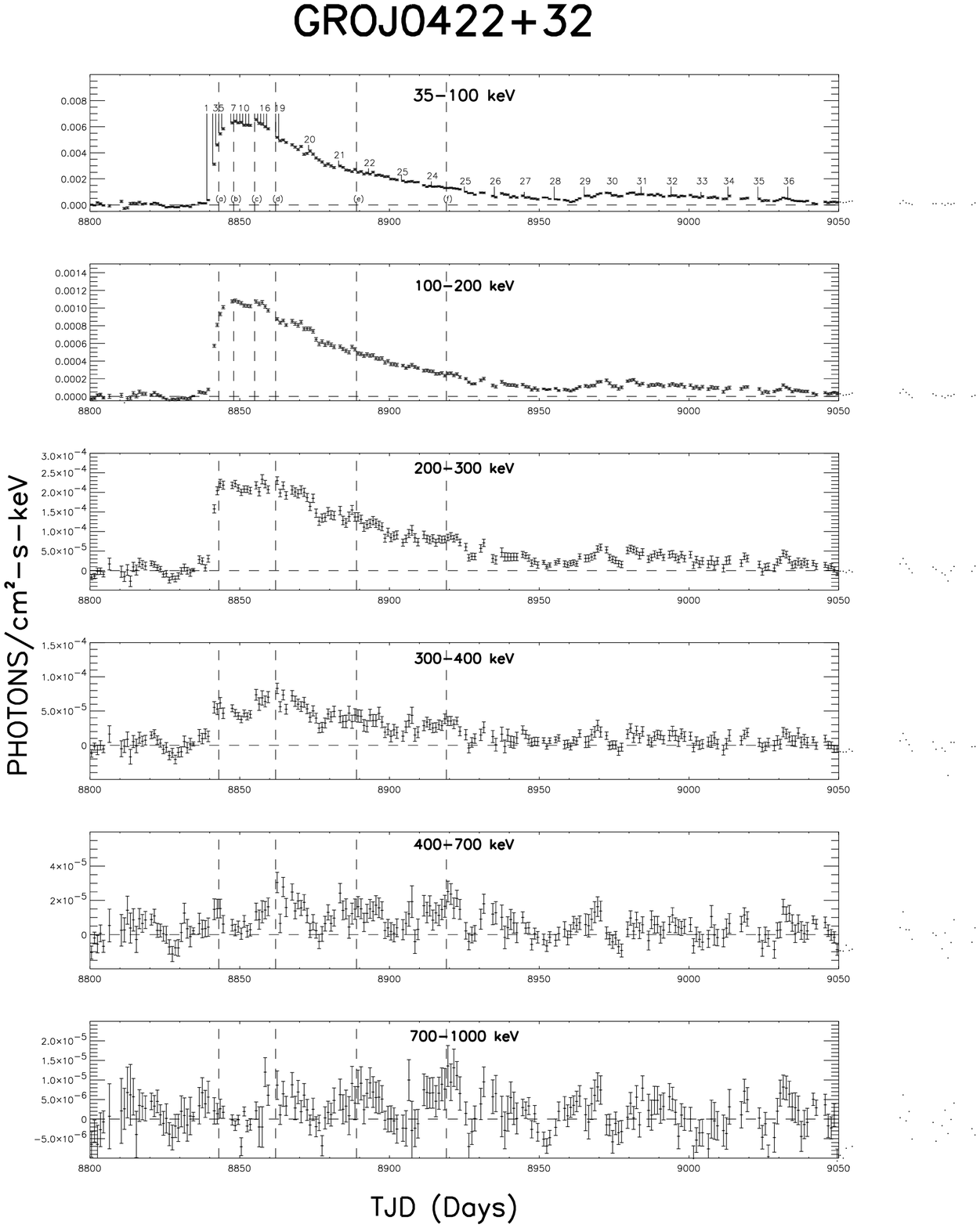}
\legend{Figure 1}
\end{figure}
\begin{figure}[t]
\centering
\includegraphics[scale=0.950]{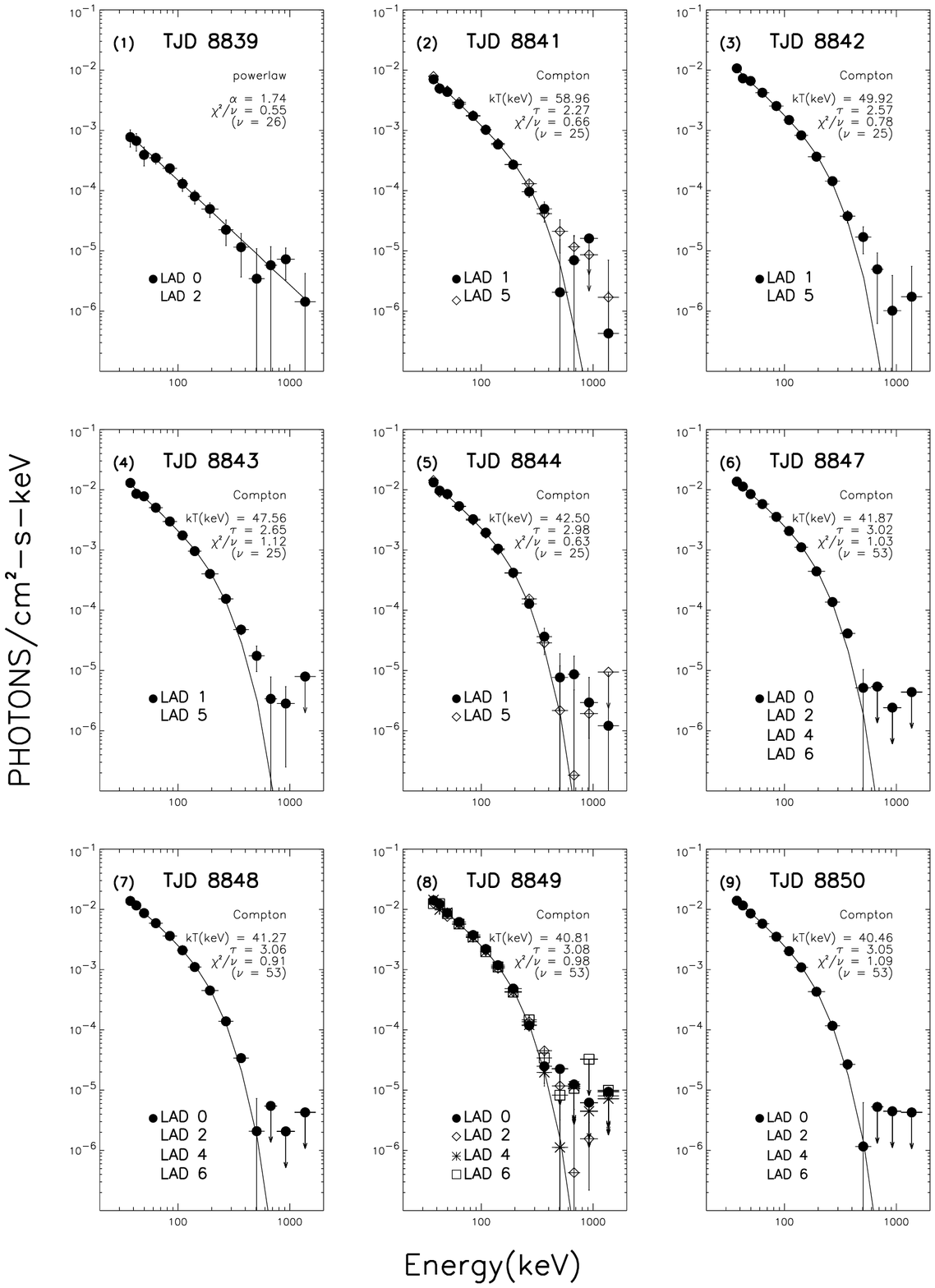}
\legend{Figure 2a}
\end{figure}
\begin{figure}[t]
\centering
\includegraphics[scale=0.950]{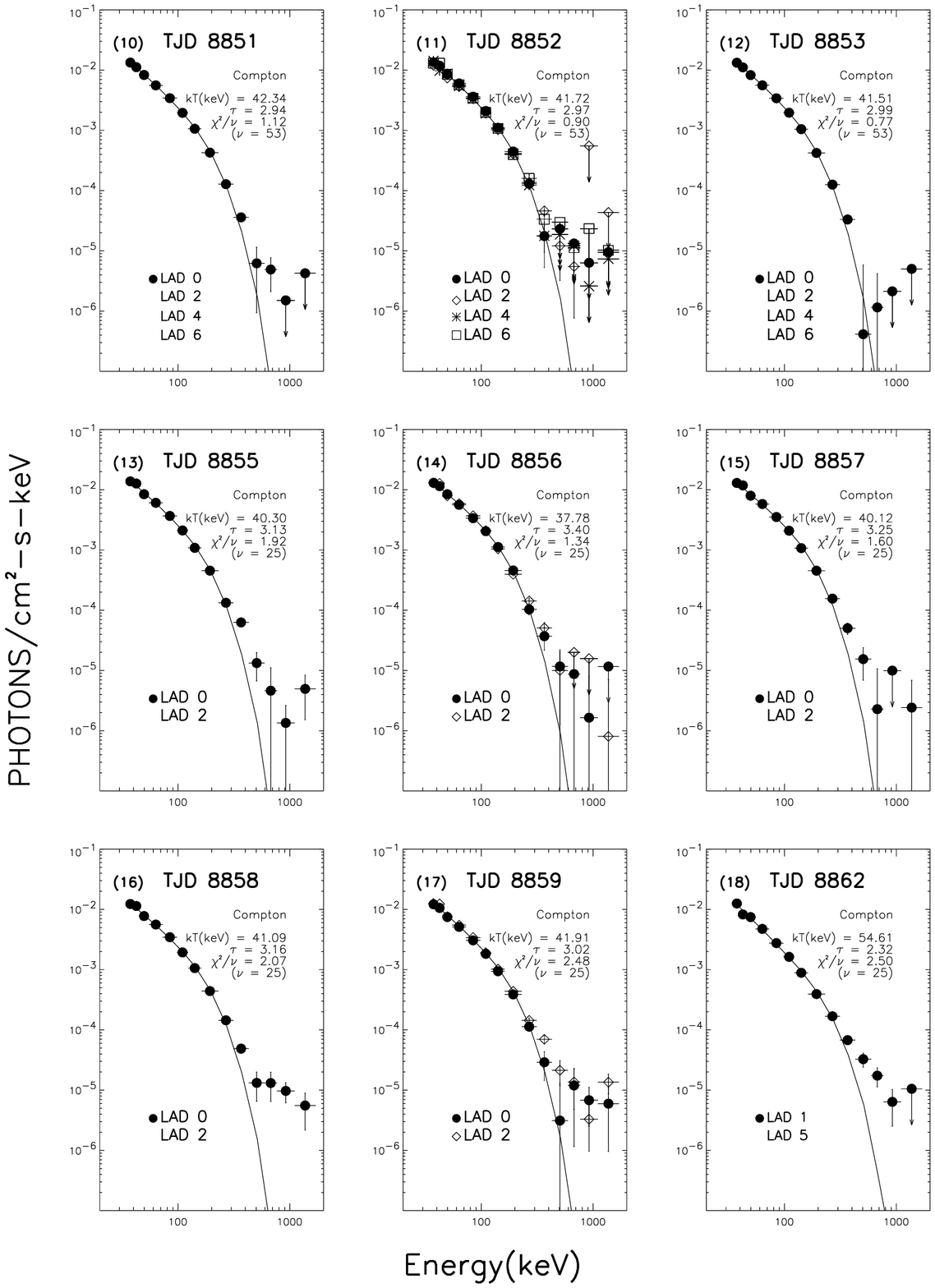}
\legend{Figure 2b}
\end{figure}
\begin{figure}[t]
\centering
\includegraphics[scale=0.950]{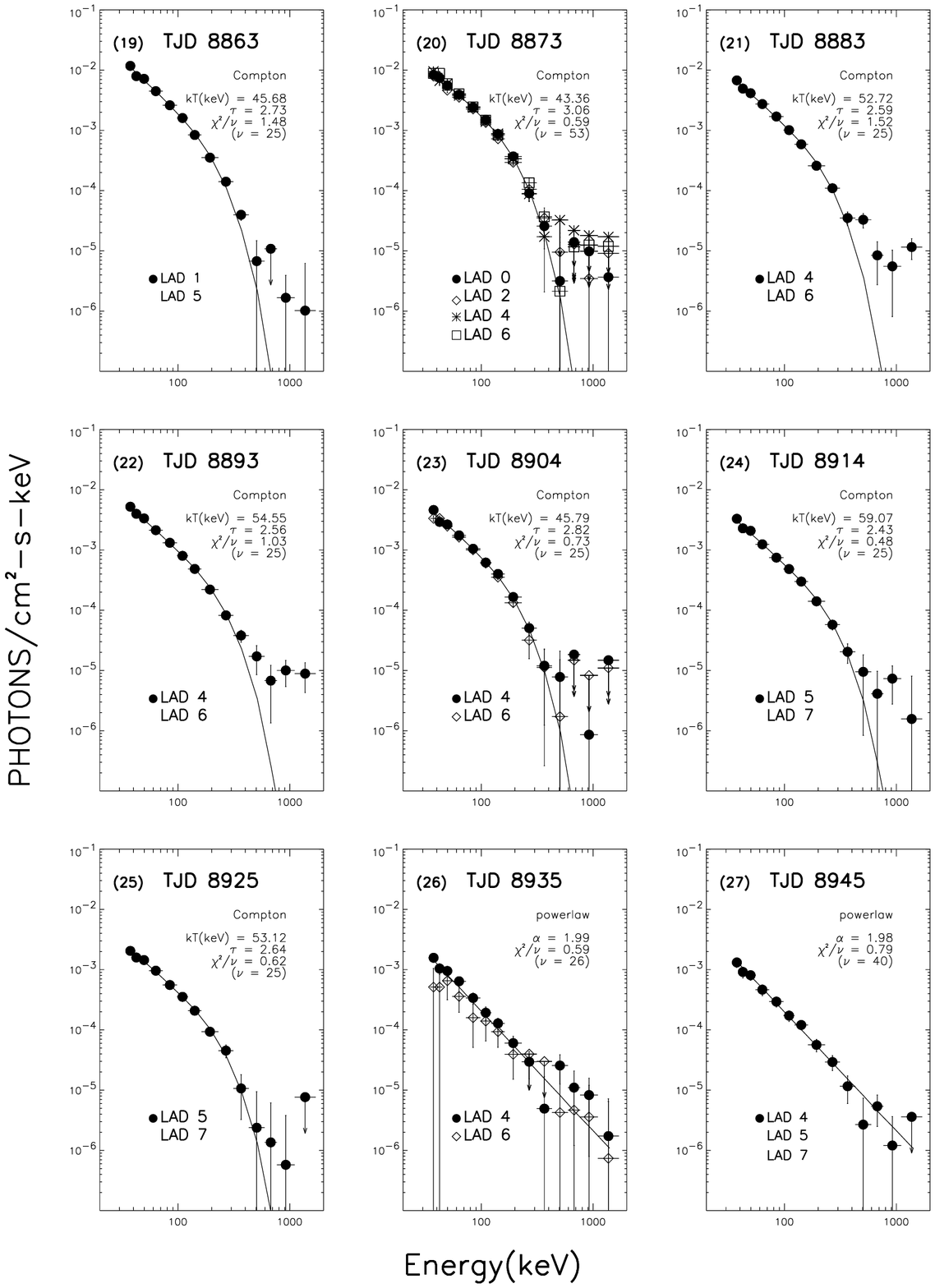}
\legend{Figure 2c}
\end{figure}
\begin{figure}[t]
\centering
\includegraphics[scale=0.950]{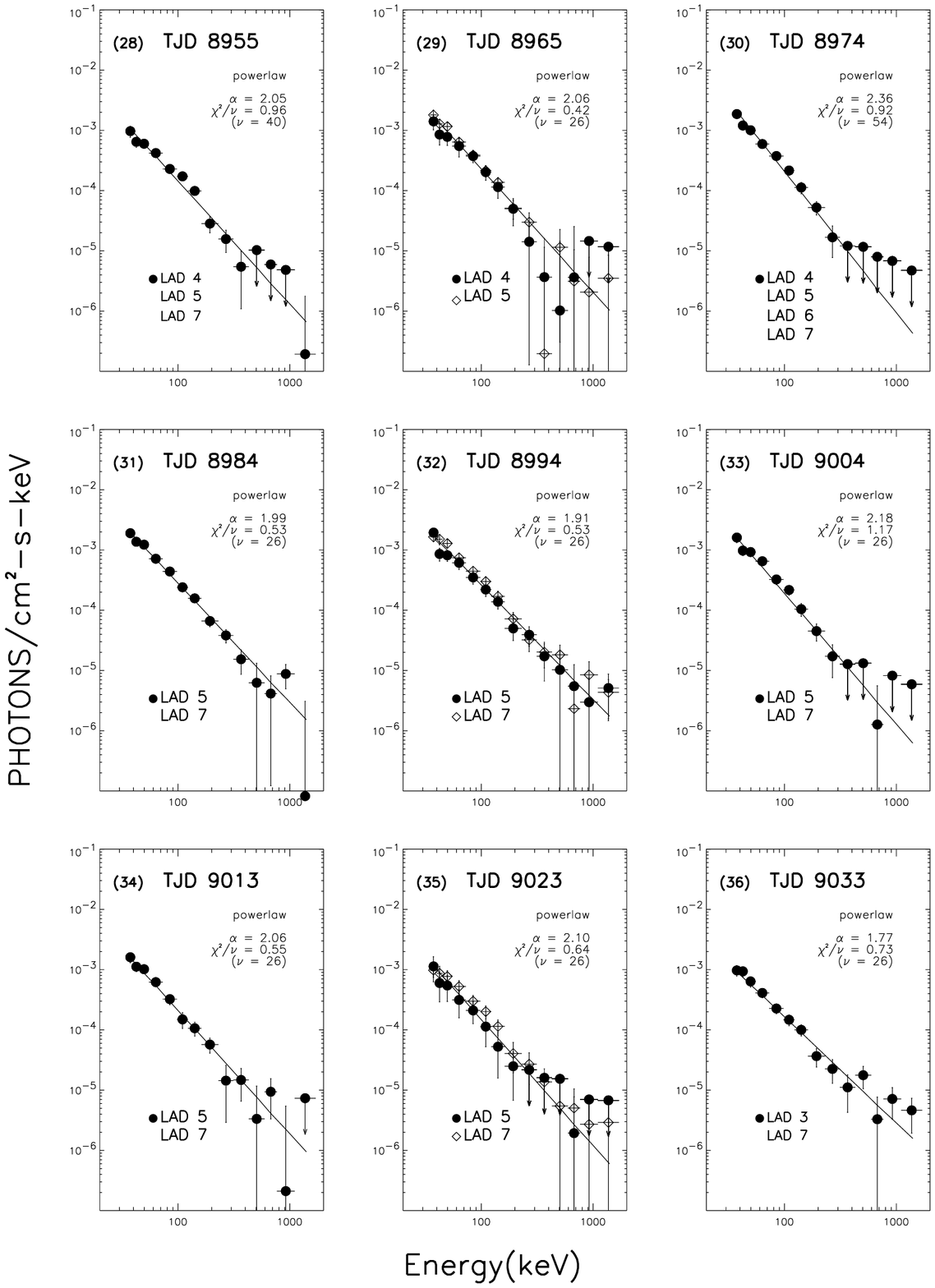}
\legend{Figure 2d}
\end{figure}
\begin{figure}[t]
\centering
\includegraphics[scale=0.950]{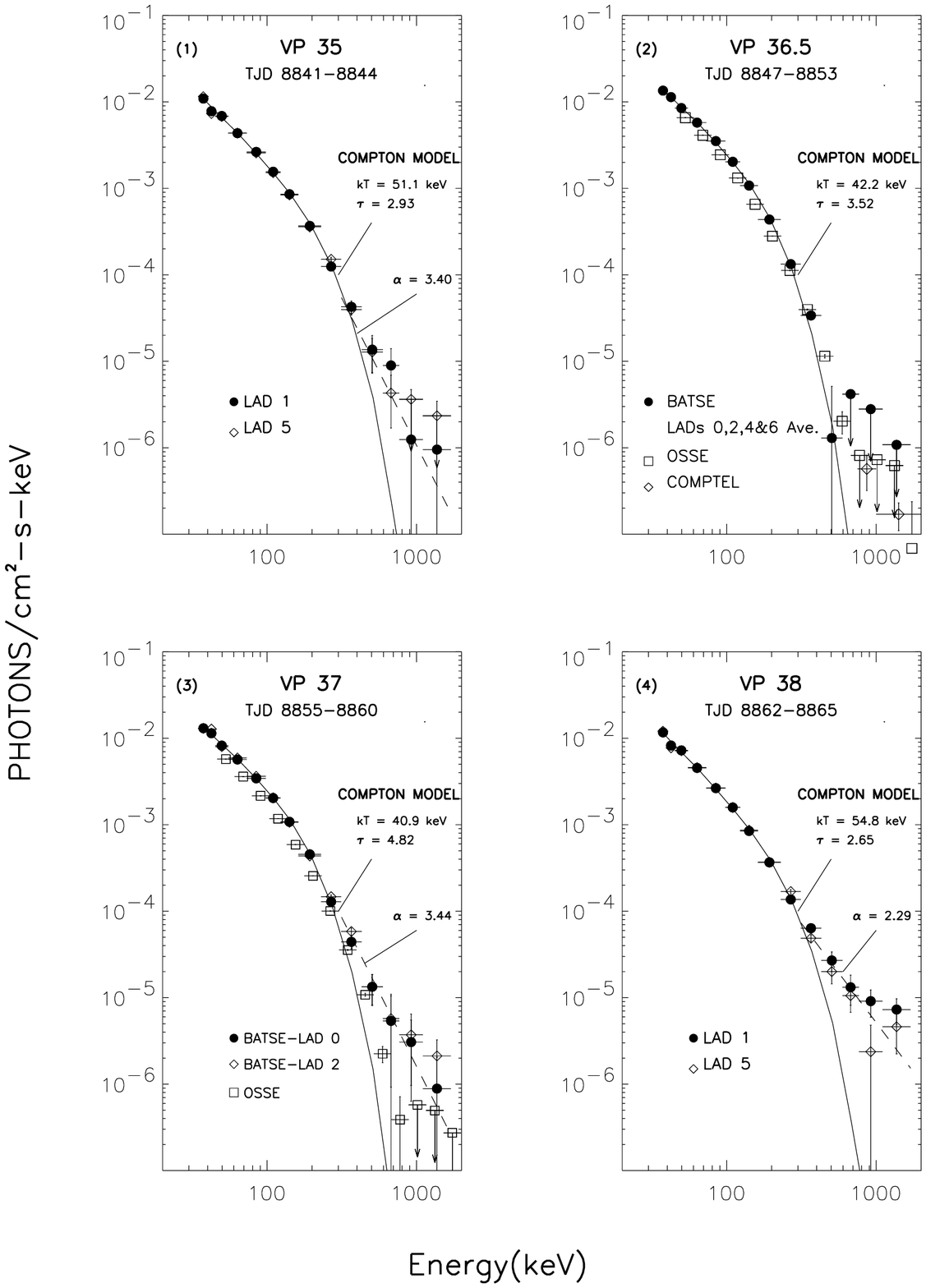}
\legend{Figure 3a}
\end{figure}
\begin{figure}[t]
\centering
\includegraphics[scale=0.950]{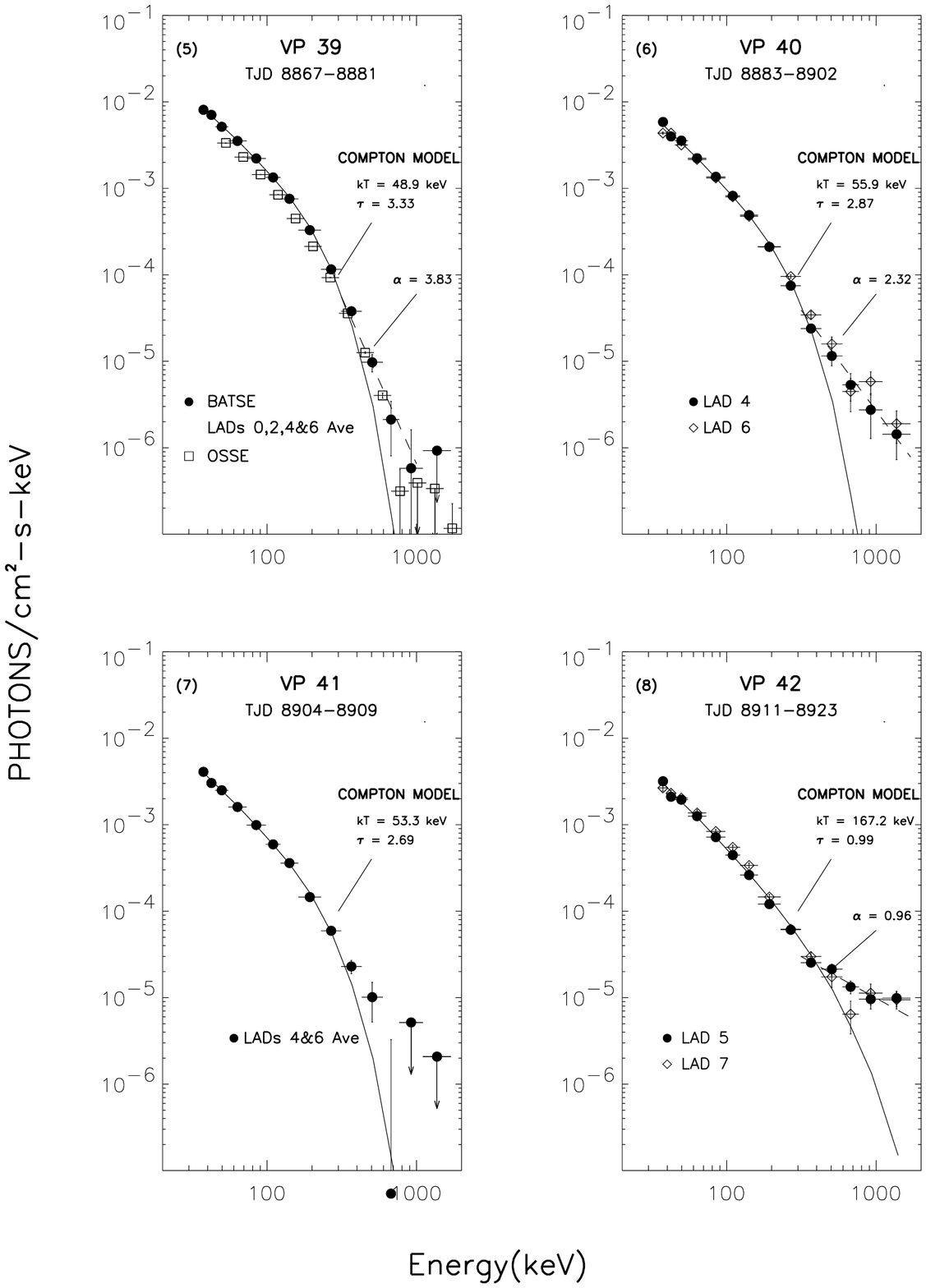}
\legend{Figure 3b}
\end{figure}
\begin{figure}[t]
\centering
\includegraphics[scale=0.950]{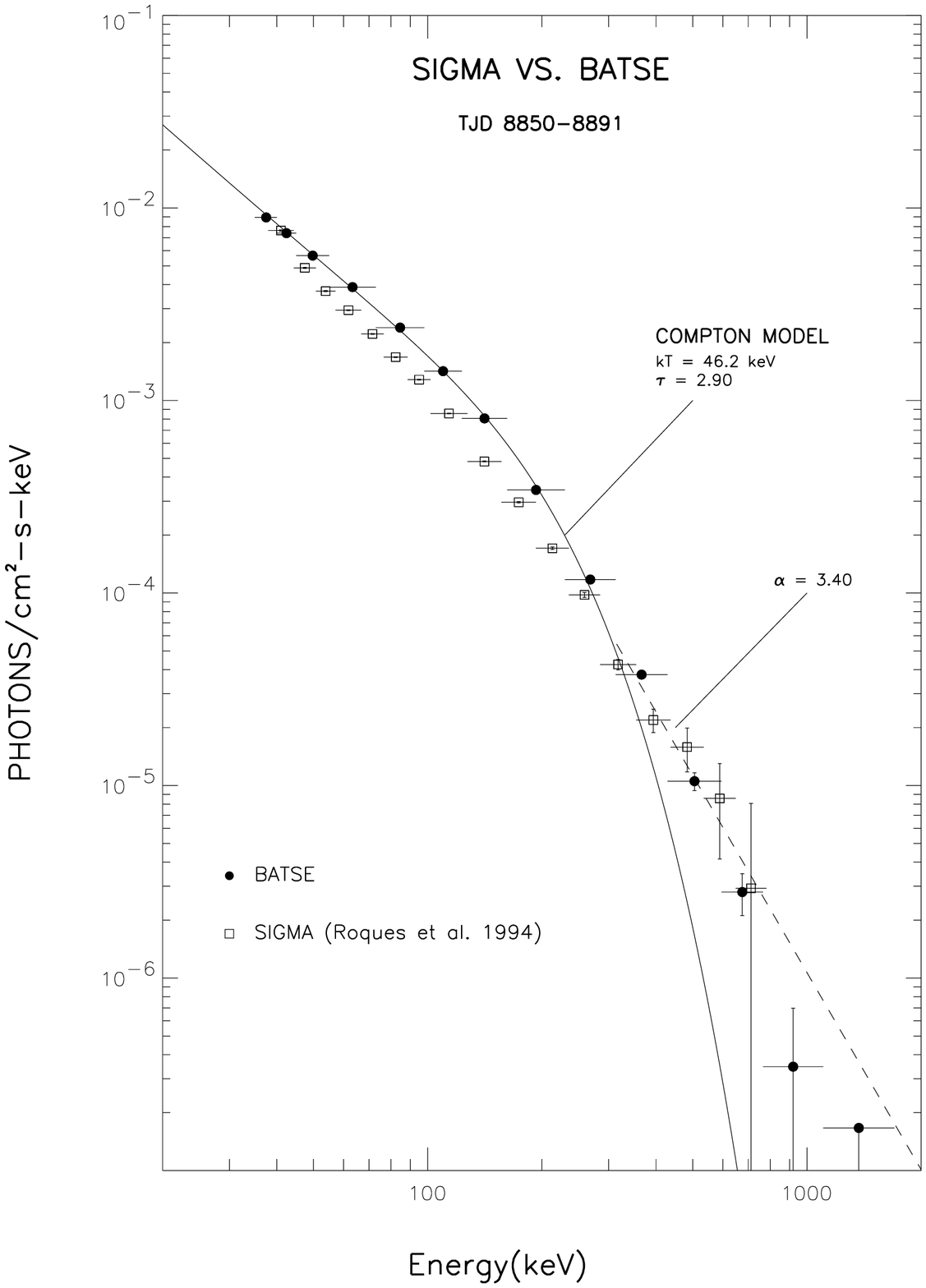}
\legend{Figure 4}
\end{figure}
\begin{figure}[t]
\centering
\includegraphics[scale=0.950]{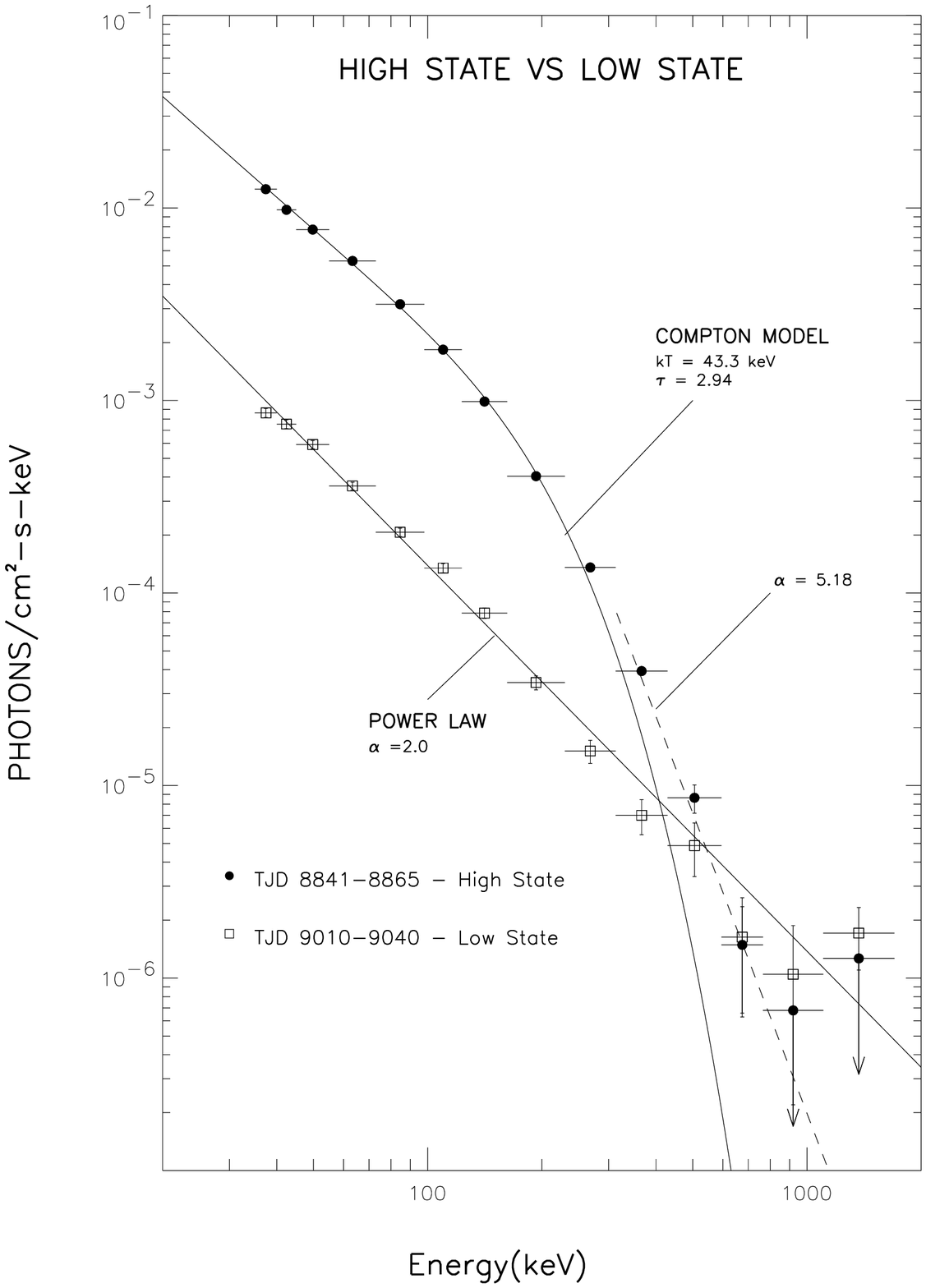}
\legend{Figure 5}
\end{figure}
\begin{figure}[t]
\centering
\includegraphics[scale=0.75]{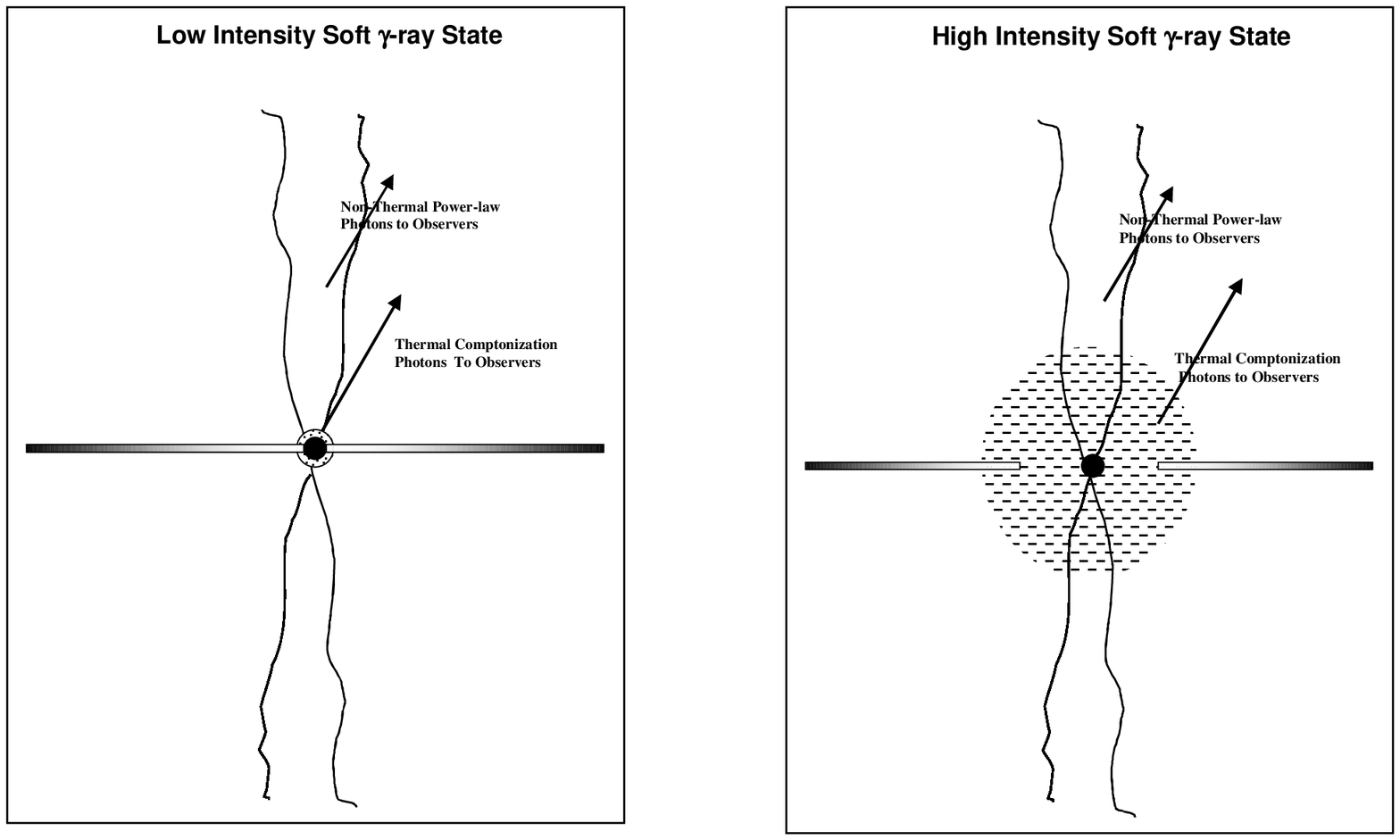}
\legend{Figure 6}
\end{figure}
\newpage
\begin{figure}[t]
\centering
\includegraphics[scale=1.0]{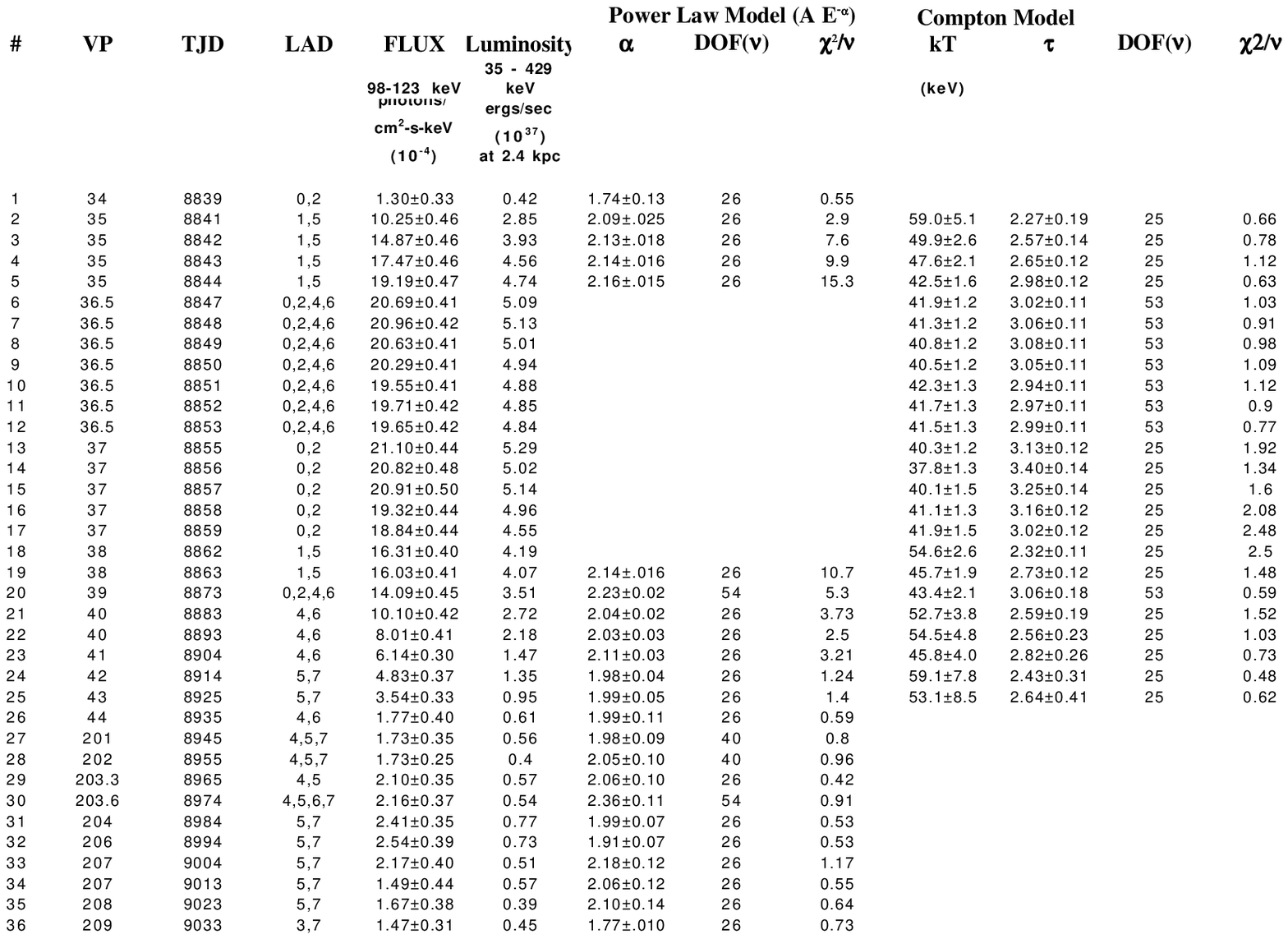}
\legend{Table 1: Best-fit parameters of power law and Compton model for the thirty-six selected single-day spectra}
\end{figure}
\newpage
\begin{figure}[t]
\centering
\includegraphics[scale=1.0]{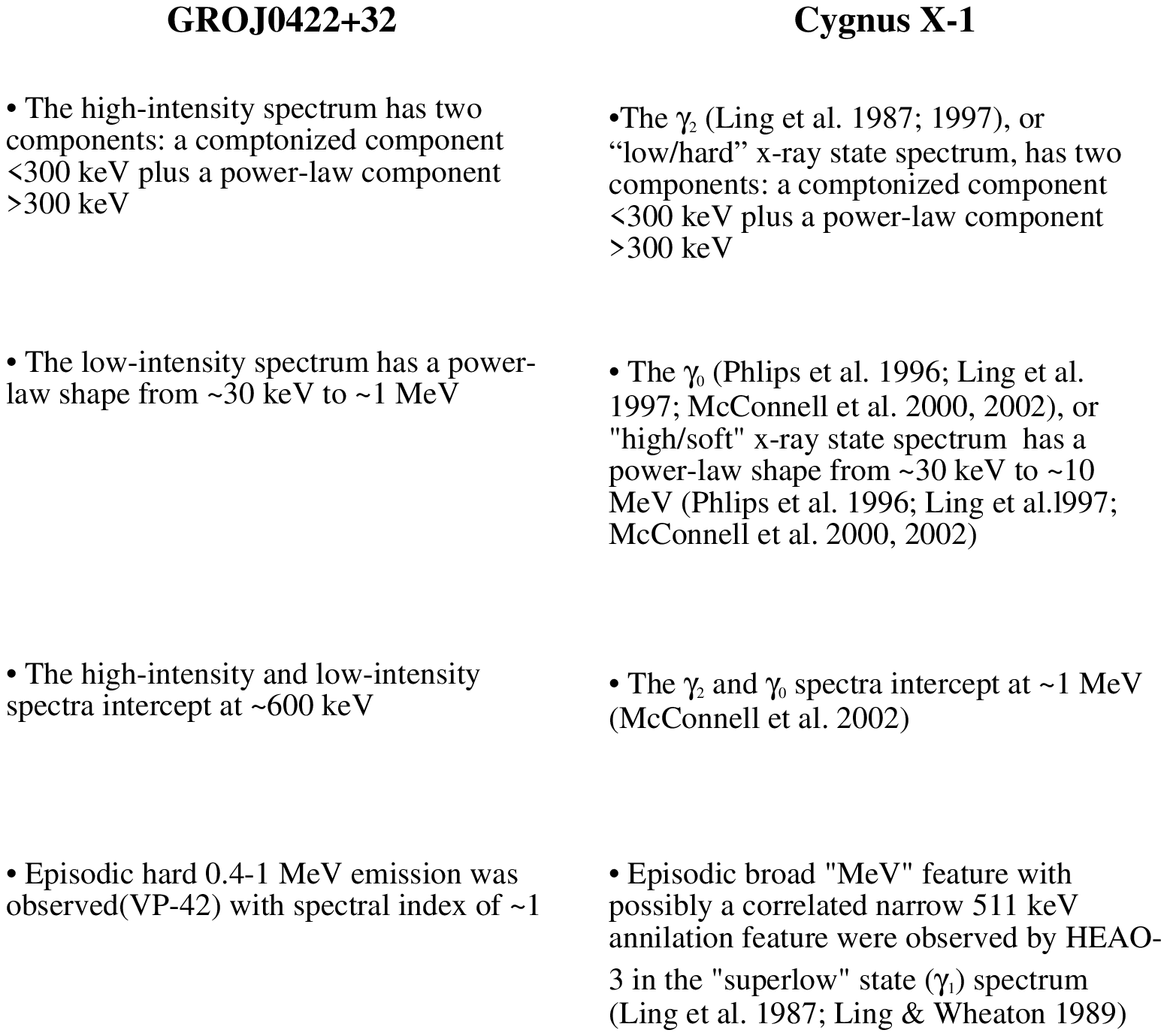}
\legend{Table 2: GROJ0422+32 \& Cygnus X-1 Comparison}
\end{figure}
\end{document}